\documentclass[fleqn,10pt]{wlscirep}
\usepackage[utf8]{inputenc}
\usepackage[T1]{fontenc}

\title{Partial Scanning Transmission Electron Microscopy with Deep Learning}

\author[1,*]{Jeffrey M. Ede}
\author[1]{Richard Beanland}
\affil[1]{University of Warwick, Department of Physics, Coventry, CV4 7AL, UK}
\affil[*]{j.m.ede@warwick.ac.uk}

\usepackage{times}
\usepackage{graphicx}
\usepackage{amssymb}
\usepackage{gensymb}
\usepackage{amsmath}
\usepackage{breakurl}

\usepackage{url,hyperref}
\hypersetup{colorlinks}

\usepackage{mathtools, cuted}

\usepackage{tabularx}
\usepackage{amsmath}

\usepackage{float}

\newcolumntype{Y}{>{\centering\arraybackslash}X}

\usepackage[]{placeins}

\newcommand\extraspace{3pt}
\newcommand\scalefiguresize{0.45}

\usepackage{pgf}

\usepackage{placeins}

\usepackage{tikz}
\newcommand*\circled[1]{\tikz[baseline=(char.base)]{
      \node[shape=circle,draw,inner sep=0.8pt] (char) {#1};}}
      
\usepackage[framemethod=tikz]{mdframed}

\begin{abstract}
Compressed sensing algorithms are used to decrease electron microscope scan time and electron beam exposure with minimal information loss. Following successful applications of deep learning to compressed sensing, we have developed a two-stage multiscale generative adversarial neural network to complete realistic 512$\times$512 scanning transmission electron micrographs from spiral, jittered gridlike, and other partial scans. For spiral scans and mean squared error based pre-training, this enables electron beam coverage to be decreased by 17.9$\times$ with a 3.8\% test set root mean squared intensity error, and by 87.0$\times$ with a 6.2\% error. Our generator networks are trained on partial scans created from a new dataset of 16227 scanning transmission electron micrographs. High performance is achieved with adaptive learning rate clipping of loss spikes and an auxiliary trainer network. Our source code, new dataset, and pre-trained models have been made publicly available at \url{https://github.com/Jeffrey-Ede/partial-STEM}.
\end{abstract}

\begin{document}

\flushbottom
\maketitle

\thispagestyle{empty}

\section{Introduction}

Aberration corrected scanning transmission electron microscopy (STEM) can achieve imaging resolutions below 0.1 nm, and locate atom columns with pm precision\cite{yankovich2015high, Peters2016}. Nonetheless, the high current density of electron probes produces radiation damage in many materials, limiting the range and type of investigations that can be performed\cite{Egerton_2004, Hujsak_2016}. A number of strategies to minimize beam damage have been proposed, including dose fractionation\cite{Jones2018} and a variety of sparse data collection methods\cite{Trampert2018}. Perhaps the most intensively investigated approach to the latter is sampling a random subset of pixels, followed by reconstruction using an inpainting algorithm\cite{anderson2013, Stevens2013, Stevens2018, Hujsak_2016, Hwang2017, Trampert2018}. Poisson random sampling of pixels is optimal for reconstruction by compressed sensing algorithms\cite{Candes2007}. However, random sampling exceeds the design parameters of standard electron beam deflection systems, and can only be performed by collecting data slowly\cite{Kovarik2016, Sang2017}, or with the addition of a fast deflection or blanking system\cite{Hujsak_2016, Beche2016}.

Sparse data collection methods that are more compatible with conventional beam deflection systems have also been investigated. For example, maintaining a linear fast scan deflection whilst using a widely-spaced slow scan axis with some small random `jitter'\cite{Kovarik2016,Stevens2018}. However, even small jumps in electron beam position can lead to a significant difference between nominal and actual beam positions in a fast scan. Such jumps can be avoided by driving functions with continuous derivatives, such as those for spiral and Lissajous scan paths\cite{Li2018, Sang2017, Sang2017a, Hujsak_2016}. Sang\cite{Sang2017, Sang2017a} considered a variety of scans including Archimedes and Fermat spirals, and scans with constant angular or linear displacements, by driving electron beam deflectors with a field-programmable gate array (FPGA) based system. Spirals with constant angular velocity place the least demand on electron beam deflectors. However, dwell times, and therefore electron dose, decreases with radius. Conversely, spirals created with with constant spatial speeds are prone to systematic image distortions due to lags in deflector responses. In practice, fixed doses are preferable as they simplify visual inspection and limit the dose dependence of STEM noise\cite{seki2018theoretical}.

Deep learning has a history of successful applications to image infilling, including image completion\cite{wu2019deep}, irregular gap infilling\cite{liu2018image} and supersampling\cite{yang2019deep}. This has motivated applications of deep learning to the completion of sparse, or `partial', scans, including supersampling of scanning electron microscopy\cite{fang2019deep} (SEM) and STEM images\cite{de2019resolution, ede2019deep}. Where pre-trained models are unavailable for transfer learning\cite{tan2018survey}, artificial neural networks (ANNs) are typically trained, validated and tested with large, carefully partitioned machine learning datasets\cite{raschka2018model, roh2019survey} so that they are robust to general use. In practice, this often requires at least a few thousand examples. Indeed, standard machine learning datasets such as CIFAR-10\cite{krizhevsky2014cifar, krizhevsky2009learning}, MNIST\cite{lecun2010mnist}, and ImageNet\cite{russakovsky2015imagenet} contain tens of thousands or millions of examples. To train an ANN to complete STEM images from partial scans, an ideal dataset might consist of a large number of pairs of partial scans and corresponding high-quality, low noise images, taken with an aberration-corrected STEM. Such a dataset does not exist. As a result, we have collated a new dataset of STEM raster scans from which partial scans can be selected. Selecting partial scans from full scans is less expensive than collecting image pairs, and individual pixels selected from experimental images have realistic noise characteristics.
 
\begin{figure}
  \centering \includegraphics[width=0.8\columnwidth]{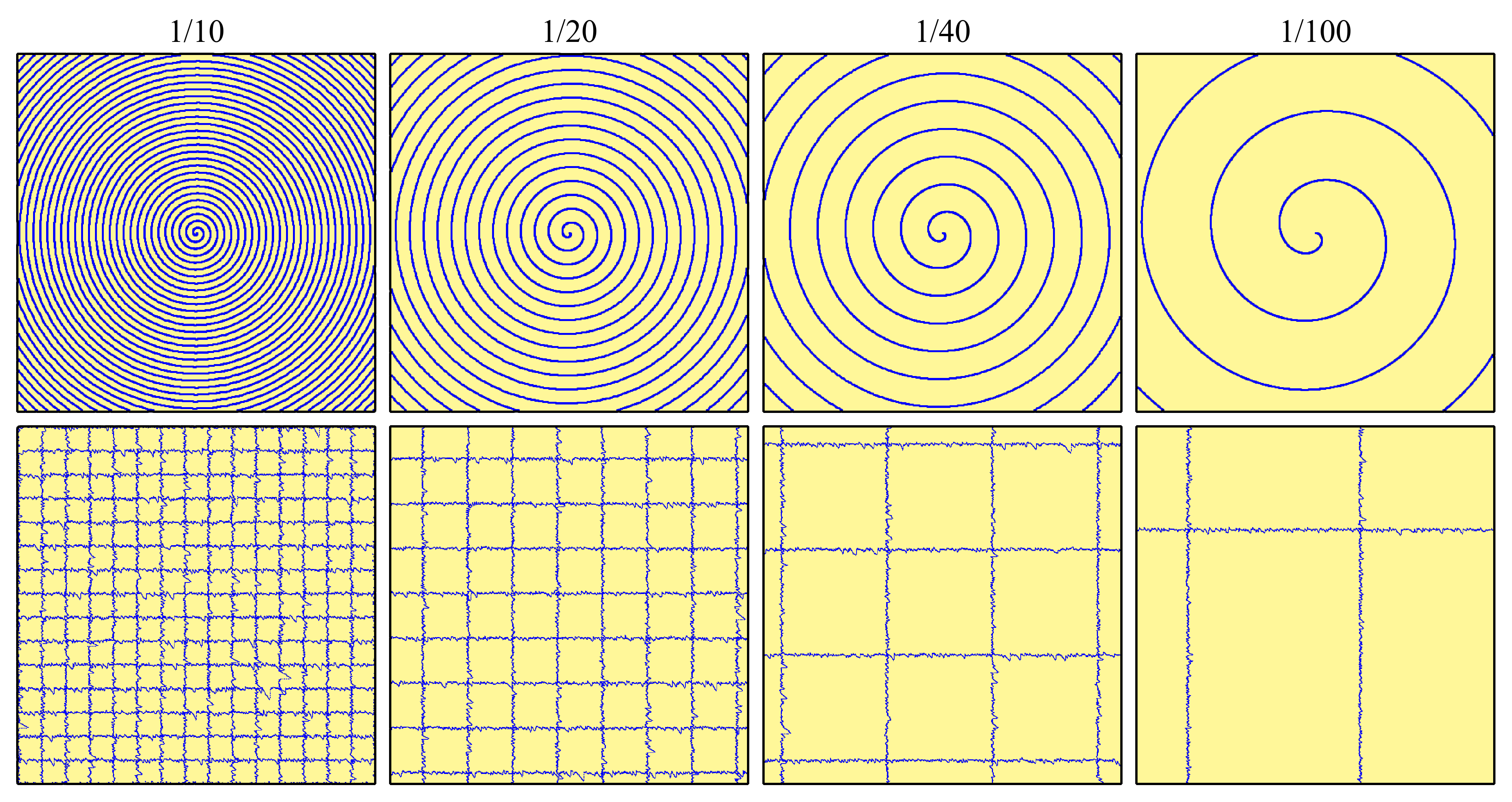}
  \caption{Examples of Archimedes spiral (top) and jittered gridlike (bottom) 512$\times$512 partial scan paths for 1/10, 1/20, 1/40, and 1/100 px coverage. }
  \label{example_walks}
\end{figure}

Examples of spiral and jittered gridlike partial scans investigated in this paper are shown in Fig.~\ref{example_walks}. Continuous spiral scan paths that extend to image corners cannot be created by conventional scan systems without going over image edges. However, such a spiral can be cropped from a spiral with radius at least $2^{-1/2}$ times the minimum image side, at the cost of increased scan time and electron beam damage to the surrounding material. We use Archimedes spirals, where $r \propto \theta$, and $r$ and $\theta$ are polar radius and angle coordinates, as these spirals have the most uniform spatial coverage. Jittered gridlike scans would also be difficult to produce with a conventional system, which would suffer variations in dose and distortions due to limited beam deflector response. Nevertheless, these idealized scan paths serve as useful inputs to demonstrate the capabilities of our approach. We expect that other scan paths could be used with similar results. 

We fine-tune our ANNs as part of generative adversarial networks\cite{goodfellow2014generative} (GANs) to complete realistic images from partial scans. A GAN consists of sets of generators and discriminators that play an adversarial game. Generators learn to produce outputs that look realistic to discriminators, while discriminators learn to distinguish between real and generated examples. Limitedly, discriminators only assess whether outputs look realistic; not if they are correct. This can result in a neural network only generating a subset of outputs, referred to as mode collapse\cite{bang2018mggan}. To counter this issue, generator learning can be conditioned on an additional distance between generated and true images\cite{mirza2014conditional}. Meaningful distances can be hand-crafted or learned automatically by considering differences between features imagined by discriminators for real and generated images\cite{wang2018high, larsen2015autoencoding}.

\section{Training}

In this section we introduce a new STEM images dataset for machine learning, describe how partial scans were selected from images in our data pipeline, and outline ANN architecture and learning policy. Detailed ANN architecture, learning policy, and experiments are provided as supplementary information, and source code is available\cite{spirals_repo}.

\subsection{Data Pipeline}

To create partial scan examples, we collated a new dataset containing 16227 32-bit floating point STEM images collected with a JEOL ARM200F atomic resolution electron microscope. Individual micrographs were saved to University of Warwick data servers by dozens of scientists working on hundreds of projects as Gatan Microscopy Suite\cite{gms_webpage} generated dm3 or dm4 files. As a result, our dataset has a diverse constitution. Atom columns are visible in two-thirds of STEM images, with most signals imaged at several times their Nyquist rates\cite{landau1967sampling}, and similar proportions of images are bright and dark field. The other third of images are at magnifications too low for atomic resolution, or are of amorphous materials. The Digital Micrograph image format is rarely used outside the microscopy community. As a result, data has been transferred to the widely supported TIFF\cite{adobe1992tiff} file format in our publicly available dataset\cite{warwickem!}.

Micrographs were split into 12170 training, 1622 validation, and 2435 test set examples. Each subset was collected by a different subset of scientists and has different characteristics. As a result, unseen validation and test sets can be used to quantify the ability of a trained network to generalize. To reduce data read times, each micrograph was split into non-overlapping 512$\times$512 sub-images, referred to as `crops', producing 110933 training, 21259 validation and 28877 test set crops. For convenience, our crops dataset is also available\cite{warwickem!}. Each crop, $I$, was processed in our data pipeline by replacing non-finite electron counts, i.e. NaN and $\pm\infty$, with zeros. Crops were then linearly transformed to have intensities $I_\text{N} \in [-1, 1]$, except for uniform crops satisfying $\max(I)-\min(I) < 10^{-6}$ where we set $I_\text{N} = 0$ everywhere. Finally, each crop was subject to a random combination of flips and 90$\degree$ rotations to augment the dataset by a factor of eight.

Partial scans, $I_\text{scan}$, were selected from raster scan crops, $I_\text{N}$, by multiplication with a binary mask $\Phi_\text{path}$,
\begin{equation}\label{eqn:noisy_scan}
I_\text{scan} = \Phi_\text{path} I_\text{N},
\end{equation}
where $\Phi_\text{path}=1$ on a scan path, and $\Phi_\text{path}=0$ otherwise. Raster scans are sampled at a rectangular lattice of discrete locations, so a subset of raster scan pixels are experimental measurements. In addition, although electron probe position error characteristics may differ for partial and raster scans, typical position errors are small\cite{ophus2016correcting, sang2014revolving}. As a result, we expect that partial scans selected from raster scans with binary masks are realistic.

We also selected partial scans with blurred masks to simulate varying dwell times and noise characteristics. These difficulties are encountered in incoherent STEM\cite{krause2016istem, hartel1996conditions}, where STEM illumination is detected by a transmission electron microscopy (TEM) camera. For simplicity, we created non-physical noise by multiplying $I_\text{scan}$ with $\eta (\Phi_\text{path}) = \Phi_\text{path} + (1 - \Phi_\text{path}) U$, where $U$ is a uniform random variate distributed in [0, 2). ANNs are able to generalize\cite{neyshabur2017exploring, kawaguchi2017generalization}, so we expect similar results for other noise characteristics. A binary mask, with values in $\{0, 1\}$, is a special case where no noise is applied i.e. $\eta (1) = 1$, and $\Phi_\text{path}=0$ is not traversed. Performance is reported for both binary and blurred masks.

The noise characteristics in our new STEM images dataset vary. This is problematic for mean squared error (MSE) based ANN training losses, as differences are higher for crops with higher noise. In effect, this would increase the importance of noisy images in the dataset, even if they are not more representative. Although adaptive ANN optimizers that divide parameter learning rates by gradient sizes\cite{ruder2016overview} can partially mitigate weighting by varying noise levels, this restricts training to a batch size of 1 and limits momentum. Consequently, we low-passed filtered ground truth images, $I_N$, to $I_\text{blur}$ by a 5$\times$5 symmetric Gaussian kernel with a 2.5 px standard deviation, to calculate MSEs for ANN outputs.

\subsection{Network Architecture}

\begin{figure}[tbp!]
\centering
\includegraphics[width=\scalefiguresize\columnwidth]{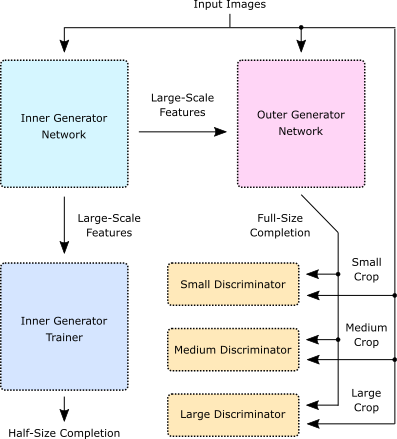}
\caption{ Simplified multiscale generative adversarial network. An inner generator produces large-scale features from inputs. These are mapped to half-size completions by a trainer network and recombined with the input to generate full-size completions by an outer generator. Multiple discriminators assess multiscale crops from input images and full-size completions. }
\label{simplified_gan}
\end{figure}

To generate realistic images, we developed a multiscale conditional GAN with TensorFlow\cite{abadi2016tensorflow}. Our network can be partitioned into the six convolutional\cite{mccann2017convolutional, krizhevsky2012imagenet} subnetworks shown in Fig.~\ref{simplified_gan}: an inner generator, $G_\text{inner}$, outer generator, $G_\text{outer}$, inner generator trainer, $T$, and small, medium and large scale discriminators, $D_1$, $D_2$ and $D_3$. We refer to the compound network $G(I_\text{scan}) = G_\text{outer}(G_\text{inner}(I_\text{scan}), I_\text{scan})$ as the generator, and to $D$ = \{$D_1$, $D_2$, $D_3$\} as the multiscale discriminator. The generator is the only network needed for inference.

Following recent work on high-resolution conditional GANs\cite{wang2018high}, we use two generator subnetworks. The inner generator produces large scale features from partial scans bilinearly downsampled from 512$\times$512 to 256$\times$256. These features are then combined with inputs embedded by the outer generator to output full-size completions. Following Inception\cite{szegedy2015going, szegedy2016rethinking}, we introduce an auxiliary trainer network that cooperates with the inner generator to output 256$\times$256 completions. This acts as a regularization mechanism, and provides a more direct path for gradients to backpropagate to the inner generator. To more efficiently utilize initial generator convolutions, partial scans selected with a binary mask are nearest neighbour infilled before being input to the generator.

Multiscale discriminators examine real and generated STEM images to predict whether they are real or generated, adapting to the generator as it learns. Each discriminator assesses different-sized crops selected from 512$\times$512 images, with sizes 70$\times$70, 140$\times$140 or 280$\times$280. After selection, crops are bilinearly downsampled to 70$\times$70 before discriminator convolutions. Typically, discriminators are applied at fractions of the full image size\cite{wang2018high} e.g. $512/2^2$, $512/2^1$ and $512/2^0$. However, we found that discriminators that downsample large fields of view to 70$\times$70 are less sensitive to high-frequency STEM noise characteristics. Processing fixed size image regions with multiple discriminators has been proposed\cite{durugkar2016generative} to decrease computation for large images, and extended to multiple region sizes\cite{wang2018high}. However, applying discriminators to arrays of non-overlapping image patches\cite{isola2017image} results in periodic artefacts\cite{wang2018high} that are often corrected by larger-scale discriminators. To avoid these artefacts and reduce computation, we apply discriminators to randomly selected regions at each spatial scale.

\subsection{Learning Policy}

Training has two halves. In the non-adversarial first half, the generator and auxiliary trainer cooperate to minimize mean squared errors (MSEs). This is followed by an optional second half of training, where the generator is fine-tuned as part of a GAN to produce realistic images. Our ANNs are trained by ADAM\cite{kingma2014adam} optimized stochastic gradient descent\cite{ruder2016overview, zou2018stochastic} for up to 2$\times$10$^6$ iterations, which takes a few days with an Nvidia GTX 1080 Ti GPU and an i7-6700 CPU. The objectives of each ANN are codified by their loss functions.

In the non-adversarial first half of training, the generator, $G$, learns to minimize the MSE based loss
\begin{equation}
L_\text{MSE} = \text{ALRC}( \lambda_\text{cond}\text{MSE}\left(G\left(I_\text{scan}), I_\text{blur}\right)\right),
\end{equation}
where $\lambda_\text{cond} = 200$, and adaptive learning rate clipping\cite{ede2019adaptive} (ALRC) limits loss spikes to stabilize learning. To compensate for varying noise levels, ground truth images were blurred by a 5$\times$5 symmetric Gaussian kernel with a 2.5 px standard deviation. In addition, the inner generator, $G_\text{inner}$, cooperates with the auxiliary trainer, $T$, to minimize 
\begin{equation}
L_\text{aux} = \text{ALRC}\left(\lambda_\text{trainer}\text{MSE}\left(T\left(G_\text{inner}\left(I_\text{scan}^\text{half}\right)\right)\right), I_\text{blur}^\text{half}\right),
\end{equation}
where $\lambda_\text{trainer} = 200$, and $I_\text{scan}^\text{half}$ and $I_\text{blur}^\text{half}$ are 256$\times$256 inputs bilinearly downsampled from $I_\text{scan}$ and $I_\text{blur}$, respectively.

In the optional adversarial second half of training, we use $N=3$ discriminator scales with numbers, $N_1$, $N_2$ and $N_3$, of discriminators, $D_1$, $D_2$ and $D_3$, respectively. There many popular GAN loss functions and regularization mechanisms\cite{wang2019generative, dong2019towards}. In this paper, we use spectral normalization\cite{miyato2018spectral} with squared difference losses\cite{mao2017least} for the discriminators, 
\begin{equation}
L_D = \frac{1}{N} \sum_{i=1}^{N} \frac{1}{N_i}[D_i(G(I_\text{scan}))^2 + (D_i(I_N) - 1)^2],
\end{equation}
where discriminators try to predict 1 for real images and 0 for generated images. We found that $N_1 = N_2 = N_3 = 1$ is sufficient to train the generator to produce realistic images. However, higher performance might be achieved with more discriminators e.g. 2 large, 8 medium and 32 small discriminators. The generator learns to minimize the adversarial squared difference loss,
\begin{equation}
L_\text{adv} = \frac{1}{N} \sum_{i=1}^{N} \frac{1}{N_i}D_i(G(I_\text{scan}) - 1)^2,
\end{equation}
by outputting completions that look realistic to discriminators.

Discriminators only assess the realism of generated images; not if they are correct. To the lift degeneracy and prevent mode collapse, we condition adversarial training on non-adversarial losses. The total generator loss is
\begin{equation}
L_G = \lambda_\text{adv} L_\text{adv} + L_\text{MSE} + \lambda_\text{aux} L_\text{aux}, 
\end{equation}
where we found that $\lambda_\text{aux}=1$ and $\lambda_\text{adv}=5$ is effective. We also tried conditioning the second half of training on differences between discriminator imagination\cite{wang2018high, larsen2015autoencoding}. However, we found that MSE guidance converges to slightly lower MSEs and similar structural similarity indexes\cite{wang2004image} for STEM images.

\begin{figure*}[tbp!]
\centering
\includegraphics[width=0.96\textwidth]{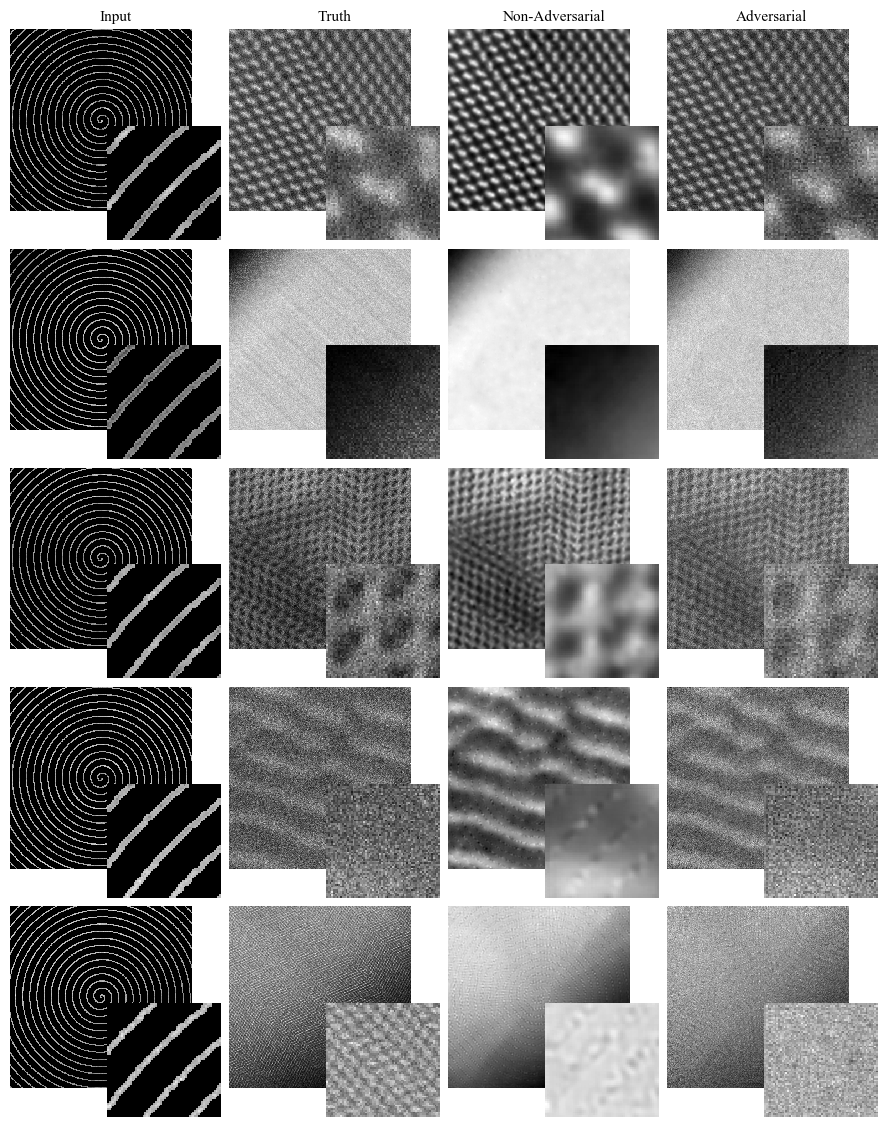}
\caption{ Adversarial and non-adversarial completions for 512$\times$512 test set 1/20 px coverage blurred spiral scan inputs. Adversarial completions have realistic noise characteristics and structure whereas non-adversarial completions are blurry. The bottom row shows a failure case where detail is too fine for the generator to resolve. Enlarged 64$\times$64 regions from the top left of each image are inset to ease comparison, and the bottom two rows show non-adversarial generators outputting more detailed features nearer scan paths. }
\label{adv_vs_non-adv}
\end{figure*}

\begin{figure*}[tbp!]
\centering
\includegraphics[width=\textwidth]{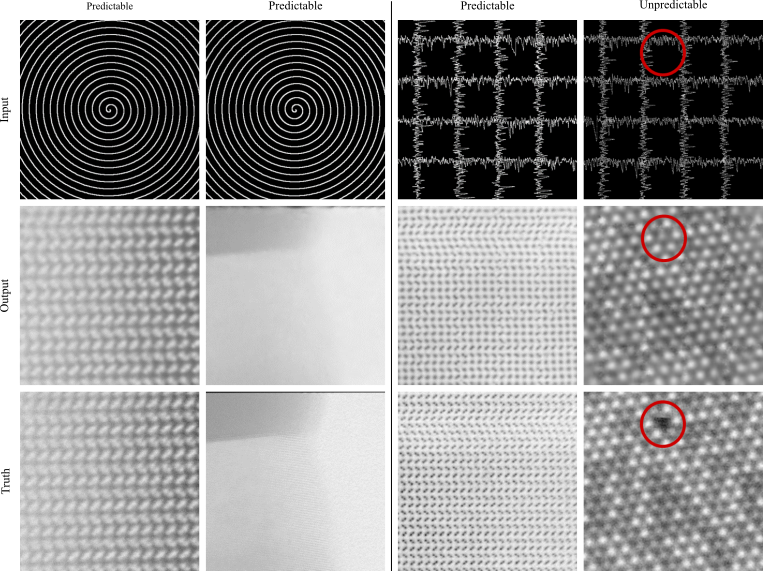}
\caption{ Non-adversarial generator outputs for 512$\times$512 1/20 px coverage blurred spiral and gridlike scan inputs. Images with predictable patterns or structure are accurately completed. Circles accentuate that generators cannot reliably complete unpredictable images where there is no information. }
\label{non-adv_example}
\end{figure*}

\begin{figure*}
\centering
\footnotesize
\includegraphics[width=\textwidth]{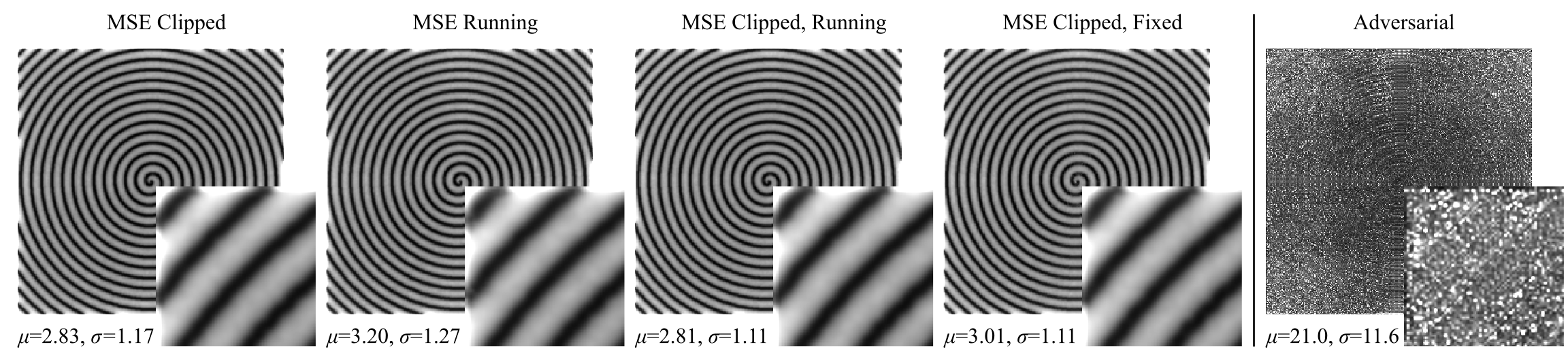}
\caption{ Generator mean squared errors (MSEs) at each output pixel for 20000 512$\times$512 1/20 px coverage test set images. Systematic errors are lower near spiral paths for variants of MSE training, and are less structured for adversarial training. Means, $\mu$, and standard deviations, $\sigma$, of all pixels in each image are much higher for adversarial outputs. Enlarged 64$\times$64 regions from the top left of each image are inset to ease comparison, and to show that systematic errors for MSE training are higher near output edges. }
\label{systematic_errors}
\end{figure*}

\begin{figure*}
\centering
\footnotesize
\includegraphics[width=\textwidth]{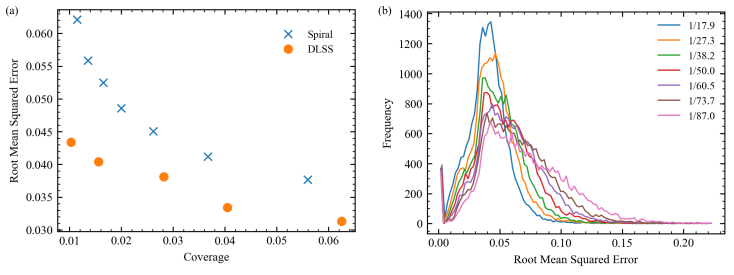}
\captionof{figure}{ Test set root mean squared (RMS) intensity errors for spiral scans in $[0, 1]$ selected with binary masks. a) RMS errors decrease with increasing electron probe coverage, and are higher than deep learning supersampling\cite{ede2019deep} (DLSS) errors. b) Frequency distributions of 20000 test set RMS errors for 100 bins in $[0, 0.224]$ and scan coverages in the legend. }
\label{err_hist}
\end{figure*}

\section{Performance}\label{sec:performance}

To showcase ANN performance, example applications of adversarial and non-adversarial generators to 1/20 px coverage partial STEM completion are shown in Fig.~\ref{adv_vs_non-adv}. Adversarial completions have more realistic high-frequency spatial information and structure, and are less blurry than non-adversarial completions. Systematic spatial variation is also less noticeable for adversarial completions. For example, higher detail along spiral paths, where errors are lower, can be seen in the bottom two rows of Fig.~\ref{adv_vs_non-adv} for non-adversarial completions. Inference only requires a generator, so inference times are the same for adversarial and non-adversarial completions. Single image inference time during training is 45 ms with an Nvidia GTX 1080 Ti GPU, which is fast enough for live partial scan completion.

In practice, 1/20 px scan coverage is sufficient to complete most spiral scans. However, generators cannot reliably complete micrographs with unpredictable structure in regions where there is no coverage. This is demonstrated by example applications of non-adversarial generators to 1/20 px coverage spiral and gridlike partial scans in Fig.~\ref{non-adv_example}. Most noticeably, a generator invents a missing atom at a gap in gridlike scan coverage. Spiral scans have lower errors than gridlike scans as spirals have smaller gaps between coverage. Additional sheets of examples for spiral scans selected with binary masks are provided for scan coverages between 1/17.9 px and 1/87.0 px as supplementary information

To characterize generator performance, MSEs for output pixels are shown in Fig.~\ref{systematic_errors}. Errors were calculated for 20000 test set 1/20 px coverage spiral scans selected with blurred masks. Errors systematically increase with increasing distance from paths for non-adversarial training, and are less structured for adversarial training. Similar to other generators\cite{ede2019improving, ede2019deep}, errors are also higher near the edges of non-adversarial outputs where there is less information. We tried various approaches to decrease non-adversarial systematic error variation by modifying loss functions. For examples: by ALRC; multiplying pixel losses by their running means; by ALRC and multiplying pixel losses by their running means; and by ALRC and multiplying pixel losses by final mean losses of a trained network. However, we found that systematic errors are similar for all variants. This is a limitation of partial STEM as information decreases with increasing distance from scan paths. Adversarial completions also exhibit systematic errors that vary with distance from spiral paths. However, spiral variation is dominated by other, less structured, spatial error variation. Errors are higher for adversarial training than for non-adversarial training as GANs complete images with realistic noise characteristics. 

Spiral path test set intensity errors are shown in Fig.~\ref{err_hist}a, and decrease with increasing coverage for binary masks. Test set errors are also presented for deep learning supersampling\cite{ede2019deep} (DLSS) as they are the only results that are directly comparable. DLSS is an alternative approach to compressed sensing where STEM images are completed from a sublattice of probing locations. Both DLSS and partial STEM results are for the same neural network architecture, learning policy and training dataset. We find that DLSS errors are lower than spiral errors at all coverages. In addition, spiral errors exponentially increase above DLSS errors at low coverages where minimum distances from spiral paths increase. Although this comparison may appear unfavourable for partial STEM, we expect that this is a limitation of training signals being imaged at several times their Nyquist rates.

Distributions of 20000 spiral path test set root mean squared (RMS) intensity errors for spiral data in Fig.~\ref{err_hist}a are shown in Fig.~\ref{err_hist}b. The coverages listed in Fig.~\ref{err_hist} are for infinite spiral paths with 1/16, 1/25, 1/36, 1/49, 1/64, 1/81, and 1/100 px coverage after paths are cut by image boundaries; changing coverage. All distributions have a similar peak near an RMS error of 0.04, suggesting that generator performance remains similar for a portion of images as coverage is varied. As coverage decreases, the portion of errors above the peak increases as generators have difficulty with more images. In addition, there is a small peak close to zero for blank or otherwise trivial completions.


\section{Discussion}\label{sec:discussion}

Partial STEM can decrease scan coverage and total electron electron dose by 10-100$\times$ with 3-6\% test set RMS errors. These errors are small compared to typical STEM noise. Decreased electron dose will enable new STEM applications to beam-sensitive materials, including organic crystals\cite{s2019low}, metal-organic frameworks\cite{mayoral2017cs}, nanotubes\cite{gnanasekaran2018quantification}, and nanoparticle dispersions\cite{ilett2019cryo}. Partial STEM can also decrease scan times in proportion to decreased coverage. This will enable increased temporal resolution of dynamic materials, including polar nanoregions in relaxor ferroelectrics\cite{kumar2019situ, xie2012static}, atom motion\cite{aydin2011tracking}, nanoparticle nucleation\cite{hussein2018tracking}, and material interface dynamics\cite{chen2018atomic}. Faster scans could also reduce delay for experimenters, decreasing microscope time.


Our generators are trained for fixed coverages and 512$\times$512 inputs. However, recent research has introduced loss function modifications that can be used to train a single generator for multiple coverages with minimal performance loss\cite{ede2019deep}. Using a single GAN improves portability as each of our GANs requires 1.3 GB of storage space with 32 bit model parameters, and limits technical debt that may accompany a large number of models. Although our generator input sizes are fixed, they can be tiled across larger images; potentially processing tiles in a single batch for computational efficiency. To reduce higher errors at the edge of generator outputs, tiles can be overlapped so that edges may be discarded\cite{ede2019improving}. Smaller images could be padded. Alternatively, dedicated generators can be trained for other output sizes.

There is an effectively infinite number of possible partial scan paths for 512$\times$512 STEM images. In this paper, we focus on spiral and gridlike partial scans. For a fixed coverage, we find that the most effective method to decrease errors is to minimize maximum distances from input information. The less information there is about an output region, the more information that needs to be extrapolated, and the higher the error. For example, we find that errors are lower for spiral scans than gridlike scans than gridlike scans as maximum distances from input information are lower. Really, the optimal scan shape is not static: It is specific to a given image and generator architecture. As a result, we are actively developing an intelligent partial scan system that adapts to inputs as they are scanned.

Partial STEM has a number of limitations relative to DLSS. For a start, partial STEM may require a custom scan system. Even if a scan system supports or can be reprogrammed to support custom scan paths, it may be insufficiently responsive. In contrast, DLSS can be applied as a postprocessing step without hardware modification. Another limitation of partial STEM is that errors increase with increasing distance from scan paths. Distances from continuous scan paths cannot be decreased without increasing coverage. Finally, most features in our new STEM crops dataset are sampled at several times their Nyquist rates. Electron microscopists often record images above minimum sufficient resolutions and intensities to ease visual inspection and limit the effects of drift\cite{jones2013identifying}, shot\cite{seki2018theoretical}, and other noise. This means that a DLSS lattice can still access most high frequency information in our dataset.

Test set DLSS errors are lower than partial STEM errors for the same architecture and learning policy. However, this is not conclusive as generators were trained for a few days; rather than until validation errors diverged from training errors. For example, we expect that spirals need more training iterations than DLSS as nearest neighbour infilled spiral regions have varying shapes, whereas infilled regions of DLSS grids are square. A lack of high frequency information also limits one of the key strengths of partial STEM that DLSS lacks: access to high-frequency information from neighbouring pixels. As a result, we expect that partial STEM performance would be higher for signals imaged closer to their Nyquist rates.

To generate realistic images, we fine-tuned partial STEM generators as part of GANs. GANs generate images with more realistic high-frequency spatial components and structure than MSE training. However, GANs focus on semantics; rather than intensity differences. This means that although adversarial completions have realistic characteristics, such as high-frequency noise, individual pixel values differ from true values. GANs can also be difficult to train\cite{salimans2016improved, liang2018generative}, and training requires additional computation. Nevertheless, inference time is the same for adversarial and non-adversarial generators after training. 

Encouragingly, ANNs are universal approximators\cite{hornik1989multilayer} that can represent\cite{lin2017does} the optimal mapping from partial scans with arbitrary accuracy. This overcomes the limitations of traditional algorithms where performance is fixed. If ANN performance is insufficient or surpassed by another method, training or development can be continued to achieve higher performance. Indeed, validation errors did not diverge from training errors during our experiments, so we are presenting lower bounds for performance. In this paper, we compare spiral STEM performance against DLSS. It is the only method that we can rigorously and quantitatively compare against as it used the same test set data. This yielded a new insight into how signals being imaged above their Nyquist rates may affect performance discussed two paragraphs earlier, and highlights the importance of standardized datasets like our new STEM images dataset. As machine learning becomes more established in the electron microscopy community, we hope that standardized datasets will also become established to standardize performance benchmarks.

Detailed neural network architecture, learning policy, experiments, and additional sheets of examples are provided as supplementary information. Further improvements might be made with AdaNet\cite{weill2019adanet}, Ludwig\cite{molino2019ludwig}, or other automatic machine learning\cite{he2019automl} algorithms, and we encourage further development. In this spirit, we have made our source code\cite{spirals_repo}, a new dataset containing 16227 STEM images\cite{warwickem!}, and pre-trained models publicly available. For convenience, new datasets containing 161069 non-overlapping 512$\times$512 crops from STEM images used for training, and 19769 antialiased 96$\times$96 area downsampled STEM images created for faster ANN development, are also available.

\section{Conclusions}

Partial STEM with deep learning can decrease electron dose and scan time by over an order of magnitude with minimal information loss. In addition, realistic STEM images can be completed by fine-tuning generators as part of a GAN. Detailed MSE characteristics are provided for multiple coverages, including MSEs per output pixel for 1/20 px coverage spiral scans. Partial STEM will enable new beam sensitive applications, so we have made our source code, new STEM dataset, pre-trained models, and details of experiments available to encourage further investigation. High performance is achieved by the introduction of an auxiliary trainer network, and adaptive learning rate clipping of high losses. We expect our results to be generalizable to SEM and other scan systems.

\section{Data Availability}

New STEM datasets are available on our publicly accessible dataserver\cite{warwickem!}. Source code is in a GitHub repository with links to pre-trained models\cite{spirals_repo}. For additional information contact the corresponding author (J.M.E.).

\bibliography{bibliography}

\section{Acknowledgements}

J.M.E. and R.B. acknowledge EPSRC grant EP/N035437/1 for financial support. In addition, J.M.E. acknowledges EPSRC Studentship 1917382.

\section{Author Contributions}

J.M.E. proposed this research, wrote the code, collated training data, performed experiments and analysis, created repositories, and co-wrote this paper. R.B. supervised and co-wrote this paper.

\section{Competing Interests}

The authors declare no competing interests.

\clearpage

\section{Detailed Architecture}\label{sec:architecture}

\begin{figure}[H]
\centering
\includegraphics[width=0.21\columnwidth]{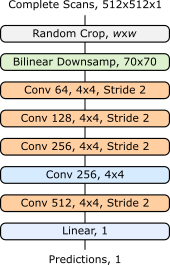}
\caption{ Discriminators examine random $w$$\times$$w$ crops to predict whether complete scans are real or generated. Generators are trained by multiple discriminators with different $w$. }
\label{discr}
\end{figure}

\begin{figure*}[tbp]
\centering
\includegraphics[width=0.97\textwidth]{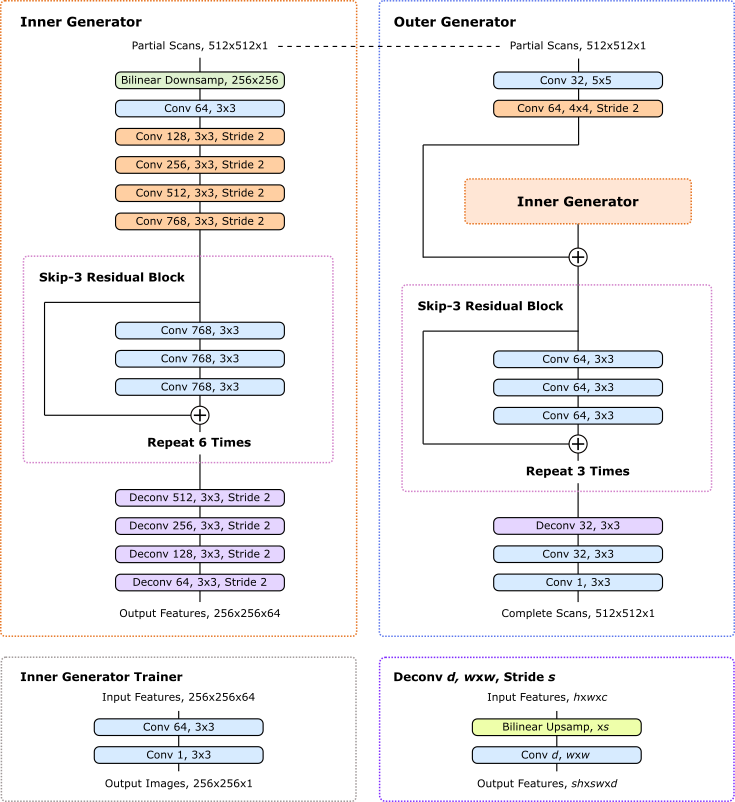}
\caption{ Two-stage generator that completes 512$\times$512 micrographs from partial scans. A dashed line indicates that the same image is input to the inner and outer generator. Large scale features developed by the inner generator are locally enhanced by the outer generator and turned into images. An auxiliary trainer network restores images from inner generator features to provide direct feedback. }
\label{gen-2-stage}
\end{figure*}

Discriminator architecture is shown in Fig.~\ref{discr}. Generator and inner generator trainer architecture is shown in Fig.~\ref{gen-2-stage}. The components in our networks are

\vspace{\extraspace}
\noindent \textbf{Bilinear Downsamp, \textit{w}x\textit{w}:} This is an extension of linear interpolation in one dimension to two dimensions. It is used to downsample images to $w$$\times$$w$.

\vspace{\extraspace}
\noindent \textbf{Bilinear Upsamp, x\textit{s}:} This is an extension of linear interpolation in one dimension to two dimensions. It is used to upsample images by a factor of $s$.

\vspace{\extraspace}
\noindent \textbf{Conv \textit{d}, \textit{w}x\textit{w}, Stride, \textit{x}:} Convolutional layer with a square kernel of width, $w$, that outputs $d$ feature channels. If the stride is specified, convolutions are only applied to every $x$th spatial element of their input, rather than to every element. Striding is not applied depthwise.

\vspace{\extraspace}
\noindent \textbf{Linear, \textit{d}:} Flatten input and fully connect it to $d$ feature channels.

\vspace{\extraspace}
\noindent \textbf{Random Crop, \textit{w}x\textit{w}:} Randomly sample a $w$$\times$$w$ spatial location using an external probability distribution. 

\vspace{\extraspace}
\noindent \textbf{\circled{+}:} Circled plus signs indicate residual connections where incoming tensors are added together. These help reduce signal attenuation and allow the network to learn perturbative transformations more easily.

All generator convolutions are followed by running mean-only batch normalization then ReLU activation, except output convolutions. All discriminator convolutions are followed by slope 0.2 leaky ReLU activation.

\section{Learning Policy}

\vspace{\extraspace}
\noindent \textbf{Optimizer}: Training is ADAM\cite{kingma2014adam} optimized and has two halves. In the first half, the generator and auxiliary trainer learn to minimize mean squared errors between their outputs and ground truth images. For the quarter of iterations, we use a constant learning rate $\eta_0 = 0.0003$ and a decay rate for the first moment of the momentum $\beta_1 = 0.9$. The learning rate is then stepwise decayed to zero in eight steps over the second quarter of iterations. Similarly, $\beta_1$ is stepwise linearly decayed to 0.5 in eight steps. In an optional second half, the generator and discriminators play an adversarial game conditioned on MSE guidance. For the third quarter of iterations, we use $\eta = 0.0001$ and $\beta_1 = 0.9$ for the generator and discriminators. In the final quarter of iterations, the generator learning rate is decayed to zero in eight steps while the discriminator learning rate remains constant. Similarly, generator and discriminator $\beta_1$ is stepwise decayed to 0.5 in eight steps.

Experiments with GAN training hyperparameters show that $\beta_1 = 0.5$ is a good choice\cite{miyato2018spectral}. Our decision to start at $\beta_1 = 0.9$ aims to improve the initial rate of convergence. In the first stage, generator and auxiliary trainer parameters are both updated once per training step. In the second stage, all parameters are updated once per training step. In most of our initial experiments with burred masks, we used a total of $10^6$ training iterations. However, we found that validation errors do not diverge if training time is increased to $2 \times 10^6$ iterations, and used this number for experiments with binary masks. These training iterations are in-line with with other GANs, which reuse datasets containing a few thousand examples for 200 epochs\cite{wang2018high}. The lack of validation divergence suggests that performance may be substantially improved, and means that our results present lower bounds for performance. All training was performed with a batch size of 1 due to the large model size needed to complete 512$\times$512 scans. 

\vspace{\extraspace}
\noindent \textbf{Adaptive learning rate clipping:} To stabilize batch size 1 training, adaptive learning rate clipping\cite{ede2019adaptive} (ALRC) was developed to limit high MSEs. ALRC layers were initialized with first raw moment $\mu_1 = 25$, second raw moment $\mu_2 = 30$, exponential decay rates $\beta_1 = \beta_2 = 0.999$, and $n = 3$ standard deviations.

\vspace{\extraspace}
\noindent\textbf{Input normalization:} Partial scans, $I_\text{scan}$, input to the generator are linearly transformed to $I_\text{scan}' = (I_\text{scan} + 1)/2$, where $I_\text{scan}' \in [0,1]$. The generator is trained to output ground truth crops in $[0,1]$, which are linearly transformed to $[-1,1]$. Generator outputs and ground truth crops in $[-1,1]$ are directly input to discriminators.

\vspace{\extraspace}
\noindent\textbf{Weight normalization:} All generator parameters are weight normalized\cite{salimans2016weight}. Running mean-only batch normalization\cite{salimans2016weight, hoffer2018norm} is applied to the output channels of every convolutional layer, except the last. Channel means are tracked by exponential moving averages with decay rates of 0.99. Running mean-only batch normalization is frozen in the second half of training to improve stability\cite{chen2017rethinking}.

\vspace{\extraspace}
\noindent \textbf{Spectral normalization:} Spectral normalization\cite{miyato2018spectral} is applied to the weights of each convolutional layer in the discriminators to limit the Lipschitz norms of the
discriminators. We use the power iteration method with one iteration per training step to enforce a spectral norm of 1 for each weight matrix. 

Spectral normalization stabilizes training, reduces susceptibility to mode collapse and is independent of rank, encouraging discriminators to use more input features to inform decisions\cite{miyato2018spectral}. In contrast, weight normalization\cite{salimans2016weight} and Wasserstein weight clipping\cite{arjovsky2017wasserstein} impose more arbitrary model distributions that may only partially match the target distribution.

\vspace{\extraspace}
\noindent\textbf{Activation:} In the generator, ReLU\cite{nair2010rectified} non-linearities are applied after running mean-only batch normalization. In the discriminators, slope 0.2 leaky ReLU\cite{maas2013rectifier} non-linearities are applied after every convolutional layer. Rectifier leakage encourages discriminators to use more features to inform decisions. Our choice of generator and discriminator non-linearities follows recent work on high-resolution conditional GANs\cite{wang2018high}.

\vspace{\extraspace}
\noindent\textbf{Initialization:} Generator weights were initialized from a normal distribution with mean 0.00 and standard deviation 0.05. To apply weight normalization, an example scan is then propagated through the network. Each layer output is divided by its L2 norm and the layer weights assigned their division by the square root of the L2 normalized output's standard deviation. There are no biases in the generator as running mean-only batch normalization would allow biases to grow unbounded c.f. batch normalization\cite{ioffe2015batch}.

Discriminator weights were initialized from a normal distribution with mean 0.00 and standard deviation 0.03. Discriminator biases were zero initialized.

\vspace{\extraspace}
\noindent \textbf{Experience replay:} To reduce destabilizing discriminator oscillations\cite{liang2018generative}, we used an experience replay\cite{pfau2016connecting, shrivastava2016learning} with 50 examples. Prioritizing the replay of difficult examples can improve learning\cite{schaul2015prioritized}, so we only replayed examples with losses in the top 20\%. Training examples had a 20\% chance to be sampled from the replay.

\section{Experiments}\label{sec:experiments}

In this section, we present learning curves for some of our non-adversarial architecture and learning policy experiments. During training, each training set example was reused $\sim$8 times. In comparison, some generative adversarial networks (GANs) are trained on the same data hundreds of times\cite{wang2018high}. As a result, we did not experience noticeable overfitting. In cases where final errors are similar; so that their difference is not significant within the error of a single experiment, we choose the lowest error approach. In practice, choices between similar errors are unlikely to have a substantial effect on performance. Each experiment took a few days with an Nvidia GTX 1080 Ti GPU. All learning curves are 2500 iteration boxcar averaged. In addition, the first $10^4$ iterations before dashed lines in figures, where losses rapidly decrease, are not shown. 

\begin{figure}[tbh!]
\centering
\includegraphics[width=0.95\textwidth]{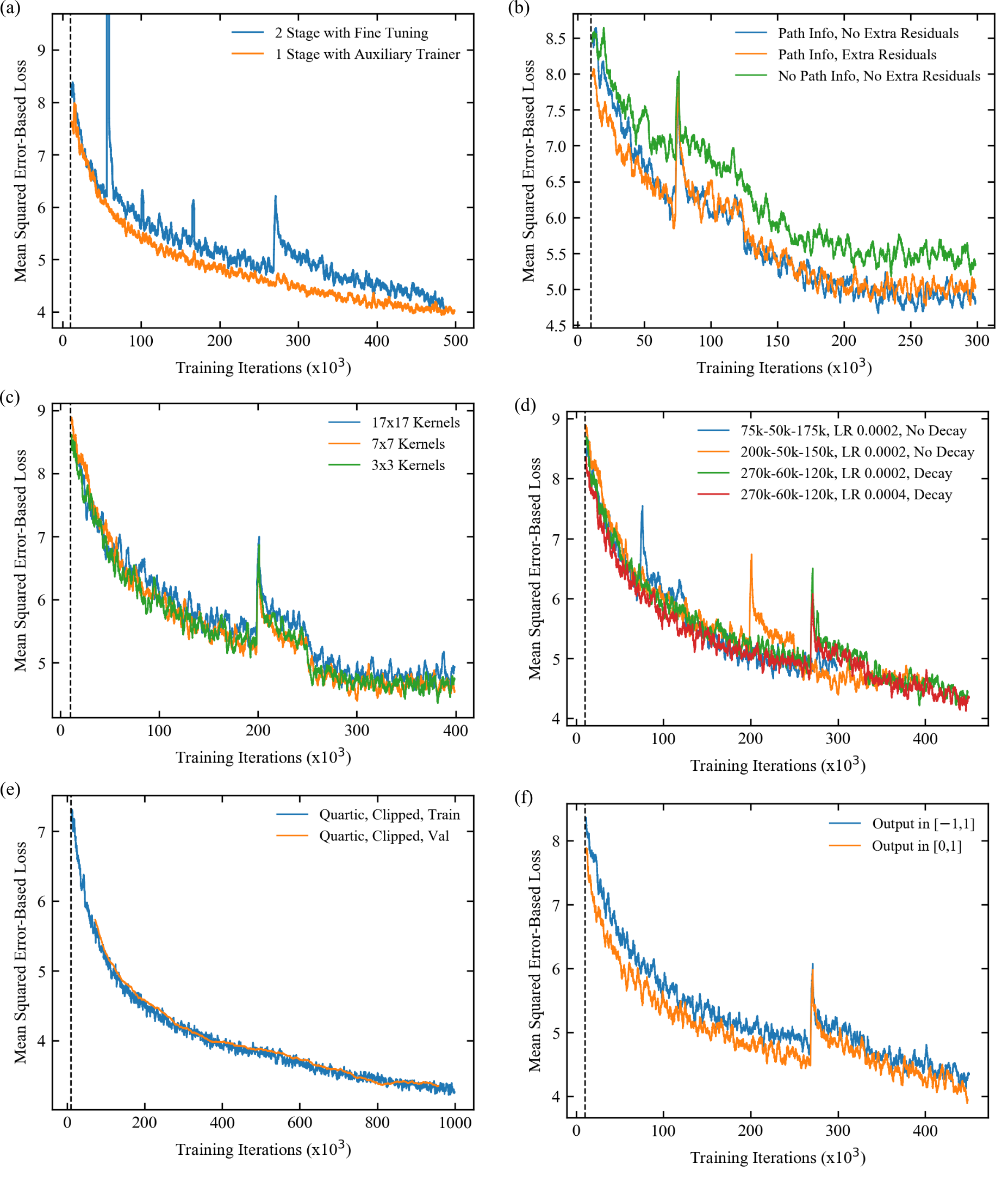}
\caption{ Learning curves. a) Training with an auxiliary inner generator trainer stabilizes training, and converges to lower than two-stage training with fine tuning. b) Concatenating beam path information to inputs decreases losses. Adding symmetric residual connections between strided inner generator convolutions and transpositional convolutions increases losses. c) Increasing sizes of the first inner and outer generator convolutional kernels does not decrease losses. d) Losses are lower after more interations, and a learning rate (LR) of 0.0004; rather than 0.0002. Labels indicate inner generator iterations - outer generator iterations - fine tuning iterations, and k denotes multiplication by 1000 e) Adaptive learning rate clipped quartic validation losses have not diverged from training losses after $10^6$ iterations. f) Losses are lower for outputs in [0, 1] than for outputs in [-1, 1] if leaky ReLU activation is applied to generator outputs. }
\label{fig:spiral_learning_curves-1}
\end{figure}

Following previous work on high-resolution GANs\cite{wang2018high}, we used a multi-stage training protocol for our initial experiments. The outer generator was trained separately; after the inner generator, before fine-tuning the inner and outer generator together. An alternative approach uses an auxiliary loss network for end-to-end training, similar to Inception\cite{szegedy2015going, szegedy2016rethinking}. This can provide a more direct path for gradients to back-propagate to the start of the network and introduces an additional regularization mechanism. Experimenting, we connected an auxiliary trainer to the inner generator and trained the network in a single stage. As shown by Fig.~\ref{fig:spiral_learning_curves-1}a, auxiliary network supported end-to-end training is more stable and converges to lower errors. 

\FloatBarrier

\begin{figure}[tbh!]
\centering
\includegraphics[width=0.95\textwidth]{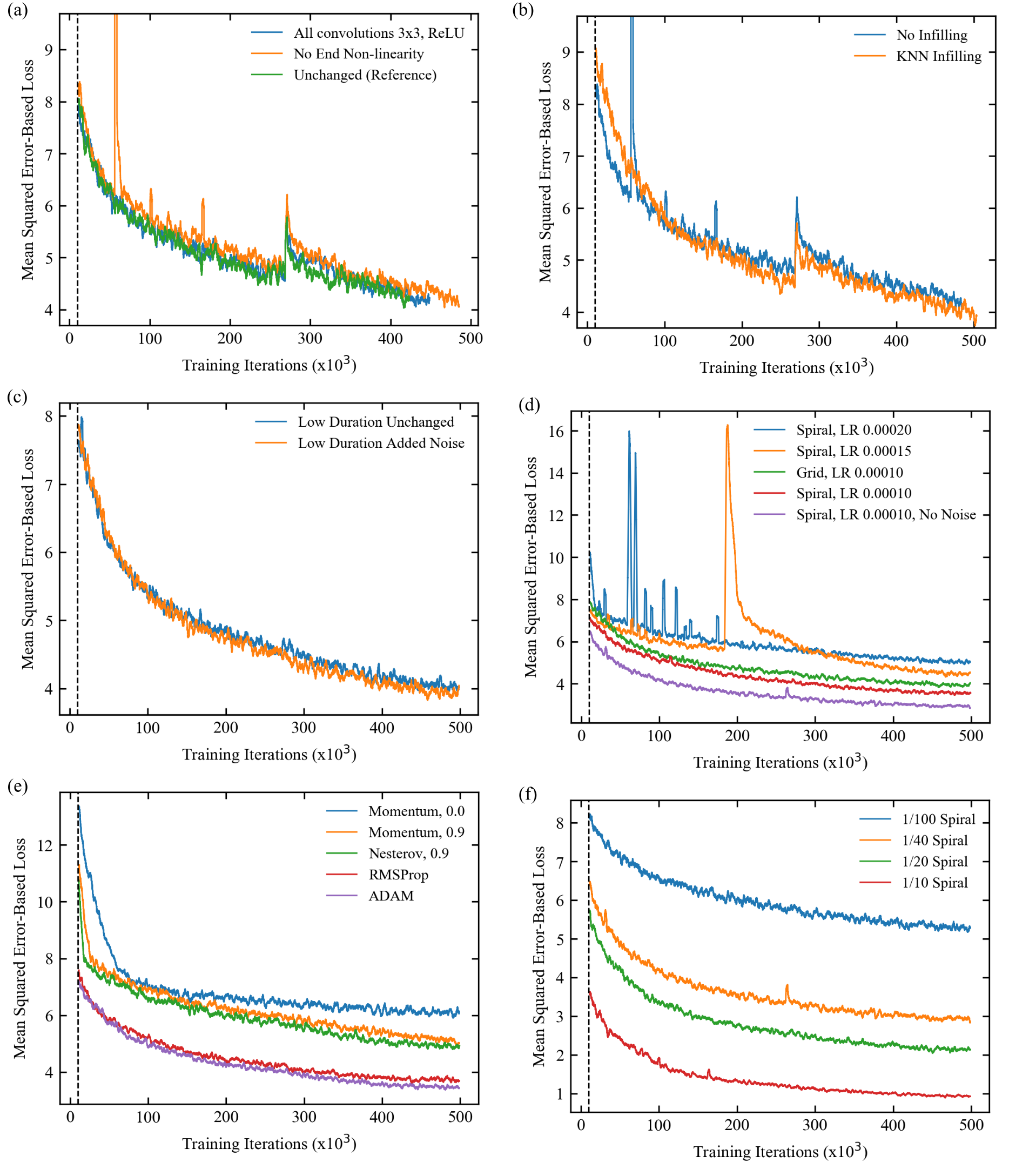}
\caption{ Learning curves. a) Making all convolutional kernels 3$\times$3, and not applying leaky ReLU activation to generator outputs does not increase losses. b) Nearest neighbour infilling decreases losses. Noise was not added to low duration path segments for this experiment. c) Losses are similar whether or not extra noise is added to low-duration path segments. d) Learning is more stable and converges to lower errors at lower learning rates (LRs). Losses are lower for spirals than grid-like paths, and lowest when no noise is added to low-intensity path segments. e) Adaptive momentum-based optimizers, ADAM and RMSProp, outperform non-adaptive momentum optimizers, including Nesterov-accelerated momentum. ADAM outperforms RMSProp; however, training hyperparameters and learning protocols were tuned for ADAM. Momentum values were 0.9. f) Increasing partial scan pixel coverages listed in the legend decreases losses. }
\label{fig:spiral_learning_curves-2}
\end{figure}

In encoder-decoders, residual connections\cite{he2016deep} between strided convolutions and symmetric strided transpositional convolutions can be used to reduce information loss. This is common in noise removal networks where the output is similar to the input\cite{mao2016image, casas2018adversarial}. However, symmetric residual connections are also used in encoder-decoder networks for semantic image segmentation\cite{badrinarayanan2017segnet} where the input and output are different. Consequently, we tried adding symmetric residual connections between strided and transpositional inner generator convolutions. As shown by Fig.~\ref{fig:spiral_learning_curves-1}b, extra residuals accelerate initial inner generator training. However, final errors are slightly higher and initial inner generator training converged to similar errors with\FloatBarrier and without symmetric residuals. Taken together, this suggests that symmetric residuals initially accelerate training by enabling the final inner generator layers to generate crude outputs though their direct connections to the first inner generator layers. However, the symmetric connections also provide a direct path for low-information outputs of the first layers to get to the final layers, obscuring the contribution of the inner generator's skip-3 residual blocks (section~\ref{sec:architecture}) and lowering performance in the final stages of training.

\begin{figure}[t!]
\centering
 \includegraphics[width=\scalefiguresize\columnwidth]{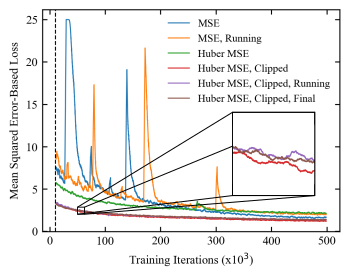}
\caption{ Adaptive learning rate clipping stabilizes learning, accelerates convergence and results in lower errors than Huberisation. Weighting pixel errors with their running or final mean errors is ineffective. }
\label{non-adversarial_14}
\end{figure}

Path information is concatenated to the partial scan input to the generator. In principle, the generator can infer electron beam paths from partial scans. However, the input signal is attenuated as it travels through the network\cite{zheng2018degeneration}. In addition, path information would have to be deduced; rather than informing calculations in the first inner generator layers, decreasing efficiency. To compensate, paths used to generate partial scans from full scans are concatenated to inputs. As shown by Fig.~\ref{fig:spiral_learning_curves-1}b, concatenating path information reduces errors throughout training. Performance might be further improved by explicitly building sparsity into the network\cite{graham2014spatially}.

Large convolutional kernels are often used at the start of neural networks to increase their receptive field. This allows their first convolutions to be used more efficiently. The receptive field can also be increased by increasing network depth, which could also enable more efficient representation of some functions\cite{lin2017does}. However, increasing network depth can also increase information loss\cite{zheng2018degeneration} and representation efficiency may not be limiting. As shown by Fig.~\ref{fig:spiral_learning_curves-1}c, errors are lower for small first convolution kernels; 3$\times$3 for the inner generator and 7$\times$7 for the outer generator or both 3$\times$3, than for large first convolution kernels; 7$\times$7 for the inner generator and 17$\times$17 for the outer generator. This suggests that the generator does not make effective use of the larger 17$\times$17 kernel receptive field and that the variability of the extra kernel parameters harms learning.

Learning curves for different learning rate schedules are shown in Fig.~\ref{fig:spiral_learning_curves-1}d. Increasing training iterations and doubling the learning rate from 0.0002 to 0.0004 lowers errors. Validation errors do not plateau for $10^6$ iterations in Fig.~\ref{fig:spiral_learning_curves-1}e, suggesting that continued training would improve performance. In our experiments, validation errors were calculated after every 50 training iterations.

The choice of output domain can affect performance. Training with a [0, 1] output domain is compared against $[-1,1]$ for slope 0.01 leaky ReLU activation after every generator convolution in Fig.~\ref{fig:spiral_learning_curves-1}f. Although $[-1,1]$ is supported by leaky ReLUs, requiring orders of magnitude differences in scale for $[-1,0)$ and $(0,1]$ hinders learning. To decrease dependence on the choice output domain, we do not apply batch normalization or activation after the last generator convolutions in our final architecture.

The $[0,1]$ outputs of Fig.~\ref{fig:spiral_learning_curves-1}f were linearly transformed to $[-1,1]$ and passed through a $\tanh{}$ non-linearity. This ensured that $[0,1]$ output errors were on the same scale as $[-1,1]$ output errors, maintaining the same effective learning rate. Initially, outputs were clipped by a tanh non-linearity to limit outputs far from the target domain from perturbing training. However, Fig.~\ref{fig:spiral_learning_curves-2}a shows that errors are similar without end non-linearites so they were removed. Fig.~\ref{fig:spiral_learning_curves-2}a also shows that replacing slope 0.01 leaky ReLUs with ReLUs and changing all kernel sizes to 3$\times$3 has little effect. Swapping to ReLUs and 3$\times$3 kernels is therefore an option to reduce computation. Nevertheless, we continue to use larger kernels throughout as we think they would usefully increase the receptive field with more stable, larger batch size training. 

To more efficiently use the first generator convolutions, we nearest neighbour infilled partial scans. As shown by Fig.~\ref{fig:spiral_learning_curves-2}b, infilling reduces error. However, infilling is expected to be of limited use for low-dose applications as scans can be noisy, making meaningful infilling difficult. Nevertheless, nearest neighbour partial scan infilling is a computationally inexpensive method to improve generator performance for high-dose applications.

To investigate our generator's ability to handle STEM noise\cite{seki2018theoretical}, we combined uniform noise with partial scans of Gaussian blurred STEM images. More noise was added to low intensity path segments and low-intensity pixels. As shown by Fig.~\ref{fig:spiral_learning_curves-2}c, ablating extra noise for low-duration path segments increases performance.

Fig.~\ref{fig:spiral_learning_curves-2}d shows that spiral path training is more stable and reaches lower errors at lower learning rates. At the same learning rate, spiral paths converge to lower errors than grid-like paths as spirals have more uniform coverage. Errors are much lower for spiral paths when both intensity- and duration-dependent noise is ablated.

To choose a training optimizer, we completed training with stochastic gradient descent, momentum, Nesterov-accelerated momentum\cite{sutskever2013importance, nesterov1983method}, RMSProp\cite{hinton2012neural} and ADAM\cite{kingma2014adam}. Learning curves are in Fig.~\ref{fig:spiral_learning_curves-2}e. Adaptive momentum optimizers, ADAM and RMSProp, outperform the non-adaptive optimizers. Non-adaptive momentum-based optimizers outperform momentumless stochastic gradient decent. ADAM slightly outperforms RMSProp; however, architecture and learning policy were tuned for ADAM. This suggests that RMSProp optimization may also be a good choice.

Learning curves for 1/10, 1/20, 1/40 and 1/100 px coverage spiral scans are shown in Fig.~\ref{fig:spiral_learning_curves-2}f. In practice, 1/20 px coverage is sufficient for most STEM images. On average, a non-adversarial generator can complete test set 1/20 px coverage partial scans with a 2.6\% root mean squared intensity error. Nevertheless, higher coverage is needed to resolve fine detail in some images. Likewise, lower coverage may be appropriate for images without fine detail. Consequently, we are developing an intelligent scan system that adjusts coverage based on micrograph content.

Training is performed with a batch size of 1 due to the large network size needed for 512$\times$512 partial scans. However, MSE training is unstable and large error spikes destabilize training. To stabilize learning, we developed adaptive learning rate clipping\cite{ede2019adaptive} (ALRC) to limit magnitudes of high losses while preserving their initial gradient distributions. ALRC is compared against MSE, Huberised MSE, and weighting each pixel's error by its Huberised running mean, and fixed final errors in Fig.~\ref{non-adversarial_14}. ALRC results in more stable training with the fastest convergence and lowest errors. Similar improvements have been confirmed  for CIFAR-10 and STEM supersampling with ALRC\cite{ede2019adaptive}.

\section{Additional Examples}\label{sec:additional_examples}

\textit{Note: Additional sheets of examples are not included in this preprint. They will be in the published version.}

Sheets of examples comparing non-adversarial generator outputs and true images are not shown in this preprint for 512$\times$512 spiral scans selected with binary masks. True images are blurred by a 5$\times$5 symmetric Gaussian kernel with a 2.5 px standard deviation so that they are the same as the images that generators were trained output. Examples are presented for 1/17.9, 1/27.3, 1/38.2, 1/50.0, 1/60.5, 1/73.7, and 1/87.0 px coverage, in that order, so that higher errors become apparent for decreasing coverage with increasing page number. Quantitative performance characteristics for each generator are provided in the main article.

\end{document}